\documentclass[conference]{IEEEtran}

\usepackage{algorithm}
\usepackage{amsmath}
\usepackage{amsfonts}
\usepackage{amssymb}
\usepackage{makeidx}
\usepackage{graphicx}
\usepackage{epstopdf}
\usepackage{cite}
\usepackage{paralist}
\usepackage{amsfonts,amssymb}
\usepackage{fancyhdr}
\usepackage{graphicx, amsfonts,amsmath, amssymb,array,url}
\usepackage{color}
\usepackage[usenames,dvipsnames,svgnames,table]{xcolor}
\usepackage{epstopdf}
\usepackage{graphicx}
\usepackage{amssymb}
\usepackage{amsmath}
\usepackage{cite}
\usepackage{subfigure}
\usepackage{mathrsfs}
\usepackage{tabulary}
\usepackage{multirow}
\usepackage[displaymath,mathlines]{lineno}
\usepackage{algpseudocode}
\usepackage{dsfont}

\ifCLASSINFOpdf

\else

\fi

\usepackage{stfloats}
\newcounter{TempEqCnt}
\IEEEoverridecommandlockouts
\begin{document}
\setlength{\columnsep}{0.22in}

\title{Downlink Interference Management \\in Dense Drone Small Cells Networks \\Using Mean-Field Game Theory}

\author{
\IEEEauthorblockN{Zihe Zhang\IEEEauthorrefmark{1},
Lixin Li\IEEEauthorrefmark{1}, Wei Liang\IEEEauthorrefmark{1},~Xu~Li\IEEEauthorrefmark{1},~Ang~Gao\IEEEauthorrefmark{1}, Wei Chen\IEEEauthorrefmark{2}, and Zhu Han\IEEEauthorrefmark{3}}
\IEEEauthorblockA{\IEEEauthorrefmark{1}School of Electronics and Information, Northwestern Polytechnical University, Xi'an, China, 710129}
\IEEEauthorblockA{\IEEEauthorrefmark{2}Tsinghua National Laboratory for Information Science and Technology (TNList)
\\Department of Electronic Engineering, Tsinghua University, Beijing, China, 100084}
\IEEEauthorblockA{\IEEEauthorrefmark{3}Department of Electrical and Computer Engineering, University of Houston, Houston, TX, USA} 
}

\maketitle

\begin{abstract}
The use of drone small cells (DSCs) has recently drawn significant attentions as one key enabler for providing air-to-ground communication services in various situations. This paper investigates the co-channel deployment of dense DSCs, which are mounted on captive unmanned aerial vehicles (UAVs). As the altitude of a DSC has a huge impact on the performance of downlink, the downlink interference control problem is mapped to an altitude control problem in this paper. All DSCs adjust their altitude to improve the available signal-to-interference-plus-noise ratio (SINR). The control problem is modeled as a mean-field game (MFG), where the cost function is designed to combine the available SINR with the cost of altitude controling. The interference introduced from a big amount of DSCs is derived through a mean-field approximation approach. Within the proposed MFG framework, the related Hamilton-Jacobi-Bellman and Fokker-Planck-Kolmogorov equations are deduced to describe and explain the control policy. The optimal altitude control policy is obtained by solving the partial differential equations with a proposed finite difference algorithm based on the upwind scheme. The simulations illustrate the optimal power controls and corresponding mean field distribution of DSCs. The numerical results also validate that the proposed control policy achieves better SINR performence of DSCs compared to the uniform control scheme.

\end{abstract}


\IEEEpeerreviewmaketitle

\vspace{0.2cm}
\section{Introduction}
\vspace{0.2cm}
\IEEEPARstart{R}{ecently}, unmanned aerial vehicles (UAVs) have become an emerging choice for air-to-ground wireless links. UAVs play an essential role in extensive application scenarios such as military environmental monitoring, damage assessments, search and so far [1], [2]. A representative UAV, loaded with wireless transmitters, has the abilities to communicate with other aerial wireless equipments as well as the equipments on the ground (referred to “users” throughout this paper). The alleged drone small cells (DSCs) [3] are mostly applied as the aerial base stations to support air-to-ground cellular link services in crucial demand and marginal situations. As a kind of highly flexible but energy-constrained devices, there are many issues on DSCs such as their management and charging. Particularly, the placement of DSCs has gained significant interests [4], [5].

In order to accommodate the evolving cellular networks which are ultra-dense and heterogeneous with dense massive equipments, the number of DSCs tends to increase. Because of the mass and distributed deployments of DSCs, the decisions of DSCs' resource allocation are anticipated to be taken individually (i.e., distributed static strategies or dynamic policies). Game theory has been proved to be an effective tool to obtain the effective distributed strategies and optimal control policies on current communication networks [6]. When facing a network of mass deployed DSCs, it is complicated for traditional game theory to settle the placement control problem due to a huge number of players. For this reason, the mean field game (MFG) [7] is an ideal tool for this sort of dense networks, where MFGs are used to model the interactions between a subjective player and the average effect of the collective behavior of other players [8], [9]. The collective behavior here can be expressed by the mean field. 

In this paper, we investigate an interference control problem under a dense DSCs downlink network, which is modelled as a MFG. As the placement of DSCs plays an essential role in the air-to-ground downlinks, each DSC can adjust the placement to improve its downlink's performance. Naturally, the competition appears resulting from the cross-tier interference, i.e., as each DSC moves to better placement to improve the quality of its downlink, it may cause higher interference to all other users served by other DSCs. The contributions of this work can be summarized as follows:
\begin{itemize}
\item	We consider the co-channel interference problem in downlink dense DSCs networks. As the altitude of a DSC is related to the distance and the elevation angle between the user to the DSC, which have a direct impact on the propagation conditions of the downlink signals, each DSC can control its vertical velocity to achieve better quality of communication by considering the interference introduced from other DSCs.

\item A MFG theoretic framework is formulated, which is ideal to model the mutual interference among a large amount of DSCs. A mean-field approximation (MFA) approach is adopted to describe the interference introduced by a mass of DSCs with the mean field. And we deduce the corresponding Hamilton-Jacobi-Bellman (HJB) and Fokker-Planck-Kolmogorov (FPK) equations for the proposed MFG framework. Since the MFG's solution can be given by handling these so-called forward and backward equations, we propose a finite difference algorithm to solve the joint differential equations. 

\item	Numerical results demonstrate the optimal power controls and corresponding mean field distribution of DSCs. We also validate the signal to noise ratio (SINR) performence of the proposed algorithm with simulations, which is promising in pratical. What's more, another advantage of this proposed algorithm is that it can be executed offline in practice.
\end{itemize}

The remainder of this paper is organized as follows. In Section \ref{section_2}, we describe the system model of co-channel dense DSCs and some assumptions. In Section \ref{section_3}, the altitude control problem is formulated to be a MFG. The forward-backward equations of this MFG is derived. In Section \ref{section_4}, we adopt a finite differential method to obtain the solutions. Simulation results are analyzed in Section \ref{section_5}. Finally, conclusions are drawn in Section \ref{section_6}.

\vspace{0.2cm}
\section{System Model}\label{section_2}
\vspace{0.2cm}
We consider a network where massive captive UAVs acting as DSCs to provide the fly wireless communications to a given geographical area. As a kind of tethered buoyant platforms, stationary captive UAVs, which are powered by carrier loaders on the ground with cable, can maintain a long time operation, so we assume that all DSCs don't have distinct horizontal placements during the considering operation period. The users on the ground can choose to access one DSC through comparing the transmit power of DSCs. Although there are more than one user connected to a DSC, we assume that only one user is served by each DSC during a given control period.

The interference problem in downlink communications is considered in this paper. Since there are DSCs using different channels in the dense DSCs networks, we concern about the interference control problem for the DSCs transmitting on the same channel. We assume that there are $N$ DSC-to-user downlinks sharing the same channel, as shown in Fig. 1. In comparison to stationary base stations, flexible DSCs in captive UAVs can choose optimal displacements to provide better downlink services to ground users. When one DSC hovering at a new altitude, which can improve the quality of its downlink, and the interference it causes to other DSCs may increase. Then the other DSCs will intuitively change their altitude.

\begin{figure}[ht]
	\begin{center}

		\includegraphics[scale=0.35]{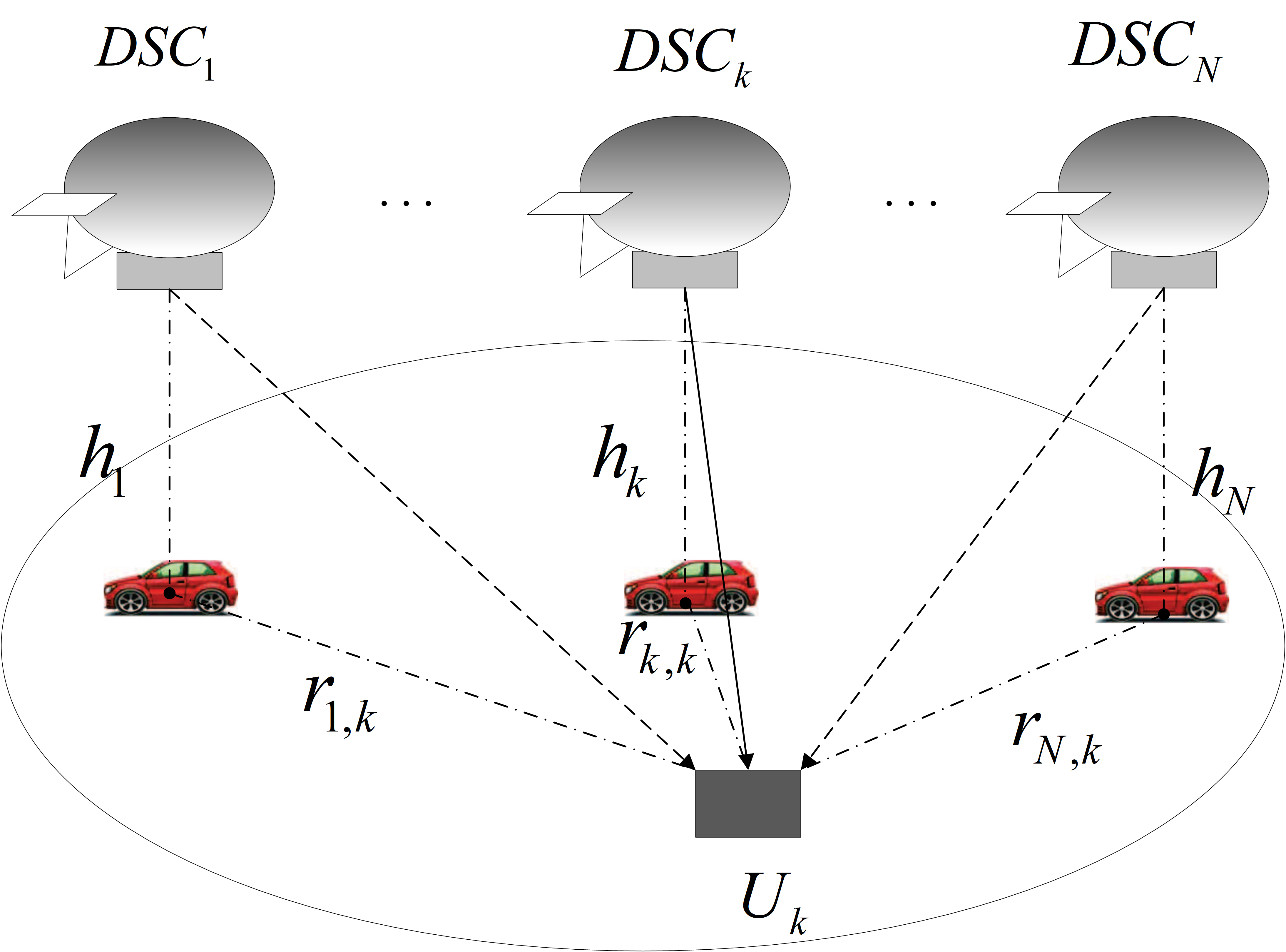}

				\caption{System model of co-channel dense DSCs.}

	\end{center}

\end{figure}

As discussed in [3], the ground users can receive three groups of signals including Line of Sight (LoS), strong reflected None Line of Sight (NLoS) signals, and multiple reflected components which cause multipath fading. We consider these signals separately with various probabilities of occurrence. We assume that the received signal is categorized in only one of those groups. Each group has its own probability of occurrence, which is a function of environment, user density, altitude of buildings and the elevation angle. Considering the LoS and NLoS connections between the user and DSC, the received signal power at each user can be expressed as

\begin{equation}
{P_{r,k,k}} = {\rm{ }}\left\{ {\begin{array}{*{20}{l}}
{{P_k}{{\left| {{X_{k,k}}} \right|}^{ - {\alpha _k}}}\;\;\;\;\;\;\;{\rm{LOS}}\;{\rm{link}},}\\
{\eta {P_k}{{\left| {{X_{k,k}}} \right|}^{ - {\alpha _k}}}\;\;\;\;\;{\rm{NLOS}}\;{\rm{link}},}
\end{array}} \right.
\end{equation}
where ${P_{r,k,k}}$ represent the $k$-th user's received signal only from the $k$-th DSC, ${P_k}$ is the transmit power of DSC $k$, $\left| {{X_{k,k}}} \right|$ is the distance between a generic user and DSC $k$, ${\alpha _k}$ is the path loss exponent over the user-DSC link, and $\eta $ is an additional attenuation factor due to the NLoS connection. And the LoS probability can be given by ${P_{LoS}} = \frac{1}{{1 + C\exp ( - B[\theta  - C])}}$, where $C$ and $B$ are fixed values which are affected by the environment (sub-urban, urban, dense urban, or others) and $\theta$, which is the elevation angle. Note that $\theta  = \frac{{180}}{\pi } \times {\sin ^{ - 1}}(\frac{h}{{\left| {{X_{k,k}}} \right|}}),$ where $\left| {{X_{k,k}}} \right| = \sqrt {{h_k}{{(t)}^2} + r_{k,k}^2} $ and ${r_{k,k}}$ is the distance between DSC $k$ and the user it serves. In our considered velocity control problem of DSCs, the transmit power of DSCs are assumed to be constant during the control period. So we denote the transmit power of DSC $k$ as $ {{P_k}}$. So the received power of the user served by DSC $k$ can be expressed as
\begin{equation}
{P_{r,k,k}} = {P_{LoS}}{P_k}{\left| {{X_{k,k}}} \right|^{ - {\alpha }}} + {P_{NLoS}}\eta {P_k}{\left| {{X_{k,k}}} \right|^{ - {\alpha }}}.
\end{equation}

The achieved SINR with the considered interferene channel at user $k$ can be presented as

\setcounter{equation}{\value{TempEqCnt}}

\setcounter{TempEqCnt}{\value{equation}}
\setcounter{equation}{9}
\begin{figure*}[hb]
\begin{equation}
u(t,h) = {\rm{E}}\int_t^T {\left( { - {\omega _1}\frac{{{P_{LoS}}{P_k}{{\left| {{X_{k,k}}} \right|}^{ - \alpha }} + {P_{NLoS}}\eta {P_k}{{\left| {{X_{k,k}}} \right|}^{ - \alpha }}}}{{N{{\bar P}_j}\int\limits_H {f\left( {m(\tau ,h),{{\bar r}_{j,k}}} \right)}  + {N_0}}} + {\omega _2}v_k^2(\tau )} \right)d\tau }.
\end{equation}
\end{figure*}
\setcounter{equation}{\value{TempEqCnt}}
\setcounter{equation}{\value{TempEqCnt}}

\setcounter{TempEqCnt}{\value{equation}}
\setcounter{equation}{15}
\begin{figure*}[ht]
\begin{equation}
{\bf{M}}(t + 1,h) = {\bf{M}}(t,h) + \frac{{\Delta t}}{{\Delta h}}\left[ {{\bf{M}}(t,h - 1){\bf{P}}(t,h - 1) - {\bf{M}}(t,h){\bf{P}}(t,h)} \right] + \frac{{{\sigma_t ^2}\Delta t}}{{2{{(\Delta h)}^2}}}\left[ {2{\bf{M}}(t,h) - {\bf{M}}(t,h + 1) - {\bf{M}}(t,h - 1)} \right].
\end{equation}
\end{figure*}
\setcounter{equation}{\value{TempEqCnt}}

\setcounter{equation}{\value{TempEqCnt}}

\setcounter{TempEqCnt}{\value{equation}}
\setcounter{equation}{17}
\begin{figure*}[ht]
\begin{equation}
\begin{array}{l}
{\bf{U}}\left( {t - 1,h} \right) = {\bf{U}}\left( {t,h} \right) + \frac{{\frac{{\Delta h}}{{\Delta t}}\left( {{\bf{U}}(t + 1,h) - {\bf{U}}(t,h)} \right) + \Delta h{\bf{C}}\left( {{{\bf{P}}^*}(t,h),{\bf{M}}(t,h)} \right)}}{{{{\bf{P}}^*}(t,h)}}\\
\;\;\;\;\;\;\;\;\;\;\;\;\;\;\;\;\;\;\; + \frac{{\sigma _t^2\Delta t}}{{2{{(\Delta h)}^2}}}\left[ {2{\bf{U}}(t,h) - {\bf{U}}(t,h + 1) - {\bf{U}}(t,h - 1)} \right].
\end{array}
\end{equation}
\end{figure*}
\setcounter{equation}{\value{TempEqCnt}}
\setcounter{equation}{2}

\begin{equation}
{\lambda _k}(t) = \frac{{{P_{r,k,k}}}}{{{I_{u,k}}(t) + {N_0}}},
\end{equation}
where ${{N_0}}$ is the thermal noise power. $ {I_{u,k}}(t) = \sum\limits_{j \in \mathbf{U},j \ne k} {{P_{r,j,k}}} $ denotes the interference introduced by other DSCs. Depending on the above definitions, the velocity control problem can be summarized as: each DSC will determine the optimal velocity control policy $ v_k^ * (t)$ considering the interference from other DSCs and the power used for displacement.

\section{Mean Field Game for Dense DSCs}\label{section_3}

In this section, the velocity control problem of dense DSCs network is modeled as a MFG. The dense DSCs network consists of $N$ DSCs, each of which can be regard as a rational policy maker in the game. The core idea of an MFG is the assumption of similarity, i.e., all players are identical and follow the same strategy [7]. They can only be differentiated by their state. If the number of players is sufficiently large, it can be assumed that the impact of a generic player on others is nearly negligible. 

\subsection{Mean Field and MFA}
In this proposed MFG, the mean field $ m(t,h) $ represents the statistical distribution of all DSCs' state, which is defined as

\begin{equation}
m(t,h) = \mathop {\lim }\limits_{K \to \infty } \frac{1}{K}\sum\limits_{\forall k \in K} {{\mathds{1}_{\left\{ {{h_k}(t) = h} \right\}}}} ,
\end{equation}
where ${\mathds{1}_{\left\{  \right\}}}$ denotes an indicator function which returns 1 if the given condition is true. Otherwise, it returns 0. For a given time instant, the mean field is the probability distribution of the states over the set of players. When the number of players $N$ is very large, we can consider that $ m(t,h) $ is a smooth continuous distribution function.  

In this MFG, we consider a optimal velocity control problem for a typical DSC. The state of the DSC is still the current altitude $h_{k}(t)$. For each DSC, it should control its velocity considering the interference introduced by all other DSCs. Benefiting from similarity, all the DSCs have the same set of equations and constraints, so the optimal control problem for the $N$ DSCs reduces to find the optimal policy for only one generic DSC. In this MFG, the infinite mass of interference is ought to be a function of the mean field when studying a typical user. To express the average interference ${{\bar I}_{u,k}}(t)$ with mean field, expect for the mean field, we need the average distance from the user served by the considered DSC to other DSCs, ${\bar r_{j,k}}$. MFA is a well-established theory in statistical physics. It essentially aims to approximate the system behavior of interacting spins in thermal equilibrium. We adopt a MFA approach to obtain approximate ${\bar r_{j,k}}$. We have the approximate expression
\begin{equation}
\begin{array}{l}
{{\bar I}_{u,k}}(t) = \sum\limits_{j = 1,j \ne k}^N {{P_{r,j,k}}}  = \sum\limits_{j = 1,j \ne k}^N {{P_j}f\left( {{h_j}(t),{r_{j,k}}} \right)}  \\
\;\;\;\;\;\;\;\;\;\;\;\;\;\;\;\;\;\;\;\;\;\;\;\;\;\;\;\;\;\;\;\;\;\;\;\;   \approx (N-1){{\bar P}_j}\bar f\left( {{h_j}(t),{r_{j,k}}} \right) ,
\end{array}
\end{equation}
where  ${\bar P_j}$ is the known test transmit power, which can be informed before the control period. And the terms ${h_j}(t),\;j \in N,j \ne k$ and ${\bar r_{j,k}}$ co-determine the mean interference channel gain as

\begin{equation}
\begin{array}{l}
f(h,{{\bar r}_{j,k}}) = \frac{1}{{1 + C\exp ( - B[{{\tan }^{ - 1}}\left( {\frac{h}{{{{\bar r}_{j,k}}}}} \right) - C])}} \cdot {\left( {{h^2} + \bar r_{j,k}^2} \right)^{ - \frac{\alpha }{2}}}\\
\;\;\; + \left( {1 - \frac{1}{{1 + C\exp ( - B[{{\tan }^{ - 1}}\left( {\frac{h}{{{{\bar r}_{j,k}}}}} \right) - C])}}} \right) \cdot \eta  \cdot {\left( {{h^2} + \bar r_{j,k}^2} \right)^{ - \frac{\alpha }{2}}}.
\end{array}
\end{equation}

We assume that the players involved in the game are using the same test transmit power and the initial state of all DSCs  ${h_j}(t),\;j \in N$ can also informed before the control period from the center controller. The term ${{\bar r}_{j,k}}$ defines the mean distance between DSC $k$'s carrier loader to other carrier loaders, which can be estimated by the following idea. If we use ${{\bar p}_j}(t)$ as the transmit power for nodes pair $i$'s transmitter, then the power received at the corresponding receiver is

\begin{equation}
P_k^r = {P_{r,k,k}} + {{\bar I}_{u,k}},
\end{equation}
where ${P_{r,k,k}}$ is the effective received power, and ${{\bar I}_{u,k}}$ is the received interference power from all the others. Thus, we can derive the only unknown variable ${{\bar r}_{j,k}}$ by above equations. As the closed-form expression of ${{\bar r}_{j,k}}$ is hard to obtain due to (6), we can apply an approximate evevation angle to reduce computational cost.

\subsection{State, Action and State equation of a Generic DSC}
We choose DSC $k$ as a generic player. The $k$-th DSC's state at time $t$ ${h_k}(t)$ is considered as the current levitated altitude. Taking into account the security issue and the limit of DSCs, the state space of DSC $k$ is constrained in $ [{h_{\min }},{h_{\max }}]$. The set of actions for DSC $k$ include all accelerated velocity, ${v_k}(t) \in \left[ {-{v_{\max }},{v_{\max }}} \right]$, where ${v_{\max }}$ is the maximum allowable velocity of any DSC. We introduce a stochastic disturbance to the state euqation considering the flapping of captive UAVs. The evolution of state for each DSC is given by following definition:

\newtheorem{myDef}{Definition}
\begin{myDef}
The state of DSC $k$ is represent by the function of time $ {h_k}(t)\in[{h_{\min }},{h_{\max }}]$ which evolves as the following differential equation:
\end{myDef}

\begin{equation}
d{h_k}(t) = {v_k}(t)dt+{\sigma _t}d{W_i}(t),\;\;\;0 \le t \le T,
\end{equation}
where we assume that the DSCs can keep a constant velocity in a short period of time through power and brake system. And the maximum absolute value of the velocity is assumed as a small value.

\subsection{Cost Function of a Generic DSC}
The communication performance of the $k$-th DSC-to-user downlink is characterized by the SINR. As the SINR is in proportionable to the probability of the successful information decoding, in our model, the DSCs tend to make their users' SINR be sufficiently large. It should be noted that a large SINR implies a low levitated altitude the DSCs will choose, which can be limited by the constrained energy. Also, a lower levitated altitude means the increase in the interference to others. Except for the energy for communication, the constrained power needs to be assigned to dynamical system for altitude shift. For simplicity, we assume the amount of power is in proportion to the quadric form of the accelarated velocity. For the altitude of the generic DSC, considering the HJB equation evolving backward in time, it should be expressed as a generic cost function, where the action (i.e., velocity) at time $t$ would only rely on the state of the DSC. The designed cost function is composed of above two elements, which can be expressed as

\begin{equation}
\begin{array}{*{20}{c}}
{{c_k}(t,h) =  - {\omega _1}\frac{{{P_{LoS}}{P_k}{{\left| {{X_{k,k}}} \right|}^{ - \alpha }} + {P_{NLoS}}\eta {P_k}{{\left| {{X_{k,k}}} \right|}^{ - \alpha }}}}{{N{{\bar P}_j}\int\limits_H {m(t,h)f\left( {h,{{\bar r}_{j,k}}} \right)} dh + {N_0}}}}\\
{ + {\omega _2}v_k^2(t,h),}
\end{array}
\end{equation}
where $ {\omega _1}$ and $ {\omega _2}$ are introduced to banlance the units of the cost elements. So each DSC should solve its optimal control problem to get the optimal accelerated velocity control policy minimizing the average cost.

\subsection{Forward-Backward Equations of MFG}

According to the Bellman’s principle of optimality, an optimal control policy should have the property that whatever the initial state and initial decision are, the remaining decisions must form an optimal policy with regard to the state resulting from the first decision. Hence, the optimal accelerated velocity control policy can then be defined in the light of a value function. The value function of the MFG is defined as (10).

The various initial states are the difference among all DSCs. Since the cost functions are only related to the mean field and the action, the MFG's HJB can be expressed as follows [9]:
\setcounter{equation}{10}
\begin{equation}
\begin{array}{*{20}{c}}
{\frac{{\partial u(t,h)}}{{\partial t}} + \mathop {\min }\limits_{v(t,h)} \left( {c(v({\rm{t}},h),m(t,h)) - v({\rm{t}},h) \cdot \frac{{\partial u(t,h)}}{{\partial h}}} \right)}\\
{ + \frac{{\sigma _t^2}}{2}\frac{{{\partial ^2}u(t,h)}}{{{\partial ^2}h}} = 0.}
\end{array}
\end{equation}

The evolution of the mean field can be described by a FPK equation. As it evolves forward in time, it is also called the forward equation. 

The forward equation of this MFG can be derived as

\begin{equation}
\frac{{\partial m(t,h)}}{{\partial t}} + \frac{\partial }{{\partial h}}\left( {m(t,h)v(t,h)} \right) + \frac{{{\sigma_t ^2}}}{2}\frac{{{\partial ^2}m(t,e)}}{{{\partial ^2}e}}= 0.
\end{equation}



To get the solution of the MFG, it is necessary to solve the two coupled PDEs derived in (11) and (12). As an FPK type equation evolves forward in time that governs the evolution of the density function of the agents. And an HJB type equation evolves backward in time that governs the computation of the optimal path for each agent.


\vspace{0.2cm}
\section{SOLUTION OF THE MFG BASED ON THE FINITE DIFFERENCE METHOD}\label{section_4}
\vspace{0.2cm}

Our goal is to obtain the identical strategy for all DSCs. As derived above, the forward-backward equations will result in the solutions of this MFG. A finite difference technique is used in this section to solve the equations. To solve the two coupled equations iteratively, the solution of the MFG can be obtained eventually.

To get the numerical solution of the MFG, the solution space is firstly discretized, where the time $T$ are discretized into a mass of intervals as $\left[ {0,{t_{\max }}\Delta t} \right]$. The step size of the time space is set as ${t_{\max }}$. Intuitively, the state of each player which satisfies the constraint $h \in [{h_{\min }},{h_{\max }}]$ can be discretized as $[{h_{\min }}\Delta h,{h_{\max }}\Delta h]$ whose step size is $\Delta h$. Accordingly, all functions respect to time and state become ${t_{\max }} \times ({h_{\max }} - {h_{\min }} + 1)$ matrices. And for notations simple, for instance, the simplified representations of the cost function at time $t\Delta t$ and state $h\Delta h$ is denoted as $C(t,h)$.  Based on the upwind difference scheme, the complicated derivative expressions with respect to time and state space in the continuous scenario can be reformulated as
\begin{equation}
\frac{{\partial U(t,h)}}{{\partial t}} = \frac{{U(t + 1,h) - U(t,h))}}{{\Delta t}},
\end{equation}
\begin{equation}
\frac{{\partial U(t,h)}}{{\partial h}} = \frac{{U(t,h) - U(t,h - 1)}}{{\Delta h}},
\end{equation}
\begin{equation}
\frac{{{\partial ^2}U(t,h)}}{{{\partial ^2}h}} = \frac{{U(t,h + 1) - 2U(t,h) + U(t,h - 1)}}{{{{(\Delta h)}^2}}}.
\end{equation}

Firstly, to evolve the mean filed, we discrete the FPK equation. Accordingly, the mean field evolution equation forward in time can be derived as (16).


The HJB equation is used to update the optimal value function. By applying the first order necessary condition on the Hamiltonian, the optimal velocity control can be derived as
\setcounter{equation}{16}
 \begin{equation}
 {v^ * }(t,h) = \frac{{\partial u(t,h)}}{{2{\omega _2}\partial h}}. 
 \end{equation}

\renewcommand{\algorithmicensure}{\textbf{Calculate:}} 
\begin{algorithm} 
\caption{Computing the solution of the MFG based on FDM.} 
\label{alg:Framwork}
\begin{algorithmic}
\Require 

Set up ${T_{\max }} \times ({h_{\max }} - {h_{\min }} + 1)$ matrices $\mathbf{U}, \mathbf{M}, \mathbf{V}$.
 
$\mathbf{M}$(0,:): Initial mean field distribution.

$\mathbf{V}(t,h)$: Arbitrary initial velocity value for the generic DSC.

$\mathbf{U}({t_{\max }},:)$ =0, i=1.

\While {$i < MAX$}

\For{all  $i = 1:1:{t_{\max }}$} 

\For{all $j \in \left\{ {{h_{\min }},...,{h_{\max }}} \right\}$} 

 Solve the FPK equation to obtain $\mathbf{M}$ with (16).
\EndFor
\EndFor

\For{all $i = {t_{\max }}:1:2$} 

\For{all $j \in \left\{ {{h_{\min }},...,{h_{\max }}} \right\}$} 

Calculate $\mathbf{U}(i-1,j)$ by using (18).

\EndFor
\EndFor

Calculate the new velocity value  ${\mathbf{V}^ * }\left( {{t_{\max }},({h_{\max }} - {h_{\min }} + 1)} \right)$ using (17).

Regressively update $\mathbf{V} = {\rm{a\mathbf{V} + b}}{{\rm{\mathbf{V}}}^ * }$ with $a + b = 1$.

\EndWhile

\end{algorithmic}
\end{algorithm}

Replacing ${v^ * }$ back to the HJB equation, the optimal value function evolution equation can be derived after some algebraic steps as (18).


To accomplish the integrated evolution of the mean field and the optimal strategy iteratively, some bounary conditions are necessary. For the computation when the time or the altitude is out of range, for instance, $\mathbf{U}(t,h + 1)$ should not exist when $h = {h_{\max }}$. So in this situation, we replace  $\mathbf{U}(t,{h_{\max }} + 1)$ by $\mathbf{U}(t,{h_{\max }})$. And to take the altitude and velocity constrains into consideration, we assume that the velocity of the DSC at the maximum and minimum altitude can't be positive and negative, respectively.

Iteratively solving (16), (17) and (18) can give a convergence solution. The whole algorithm for getting the solution of the MFG is given in Algorithm \ref{alg:Framwork}. To operate this algorithm, each DSC should know the initial mean field distributions, which can be informed from the center controller. And to express the cost function, ${\bar r_{j,k}}$ is obtained based on the MFA before the beginning of the game. The algorithm stops when the number of iterations is larger than a threshold value.

\begin{figure}[ht]
	\begin{center}

		\includegraphics[scale=0.30]{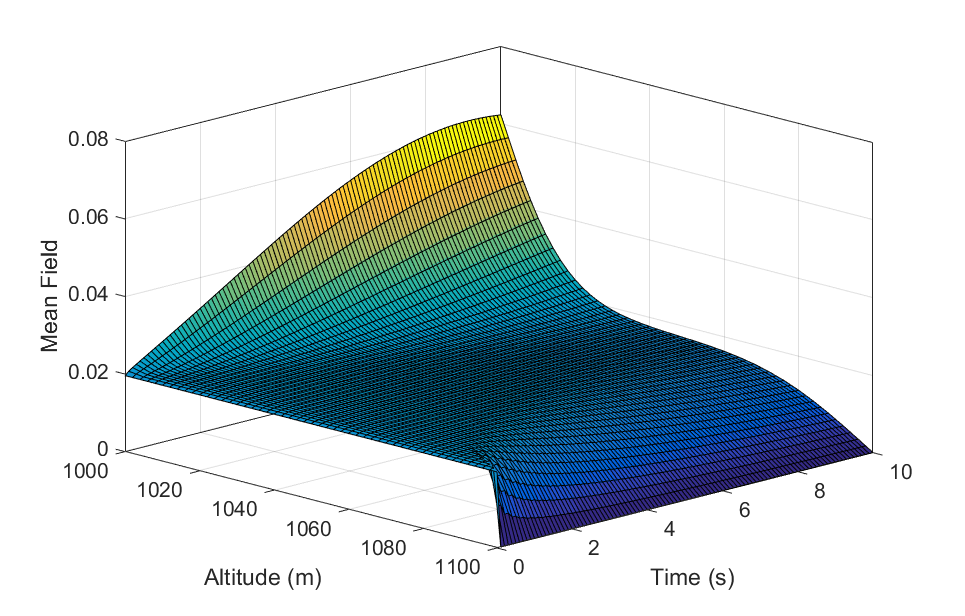}

				\caption{Mean field distribution with a uniform initial altitude distribution.}

	\end{center}

\end{figure} 
\begin{figure}[ht]
	\begin{center}

		\includegraphics[scale=0.30]{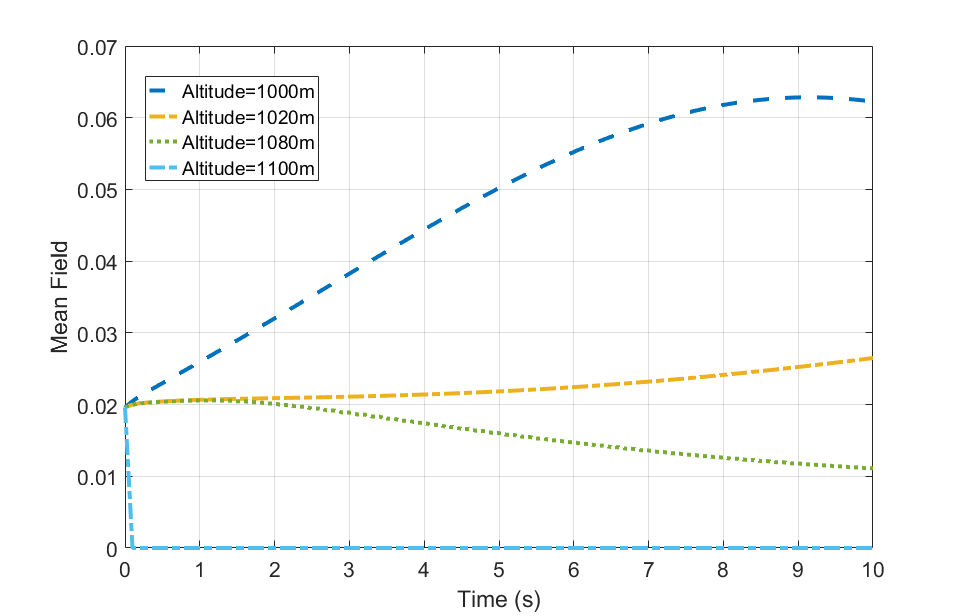}

				\caption{Cross-sections of the mean field distributions.}

	\end{center}

\end{figure}

\section{Numerical Results and Discussion}\label{section_5}

In this section, the system behaviors and the distributions of the mean field are diagrammatized. Note that the captive UAVs in this MFG opearte as Low Altitude Platforms (LAPs), which are hovering at a low altitude range. We set  ${h_{\min }}$ =1000m and ${h_{\max }}$ =1100m, $\left| {{v_{\max }}} \right|$ =10m/s, $f=2000MHz$ and $T = 10s$. The transmit power is set as 10$dB$, $\alpha$=2 and $\eta$=0.05. And we focus on the urban environment, whose environment parameters  $C = 10$ and $B = 0.03$, which is calculated with the algorithm in [10].

Fig. 2 illustrates the mean field distributions based on the optimal velocity controls of DSCs. We can observe that the number of higher DSCs decreases with time. And the DSCs that hovering at lower altitudes gradually increase with time. This implies that all DSCs tend to slew down after the start. Although there is an optimal elevation angle for each DSC, which can give the maximum probability of having LoS connections [3], the decrease of the altitude can more significantly reduce the path loss. 
 
As shown in Fig. 3, some cross-sections of the from the mean field give a clearer view to the variation of the probability distribution of DSCs hovering at a certain altitude with time. In this simulation, the initial probability are identical since we assume the mean field is a uniform distribution at the beginning. The goal of each DSC is to maximize the SINR. Therefore, the probability of DSCs hovering at the maximum altitude dives to none after the beginning. And the number of DSCs hovering at the minimum altitude increases with time. The probability distribution of DSCs at 1020m has a slight increase at last. The mean field at 1080m keeps stable for a while, and decreases after that. This is because that after some time, there are few DSCs at higher altitudes. There is another phenomenon that the rates of the variations of these cross-sections decrease with time. The reason is that along with time, more DSCs converge at lower altitudes, 
 \begin{figure}[ht]
 	\begin{center}
 
 		\includegraphics[scale=0.30]{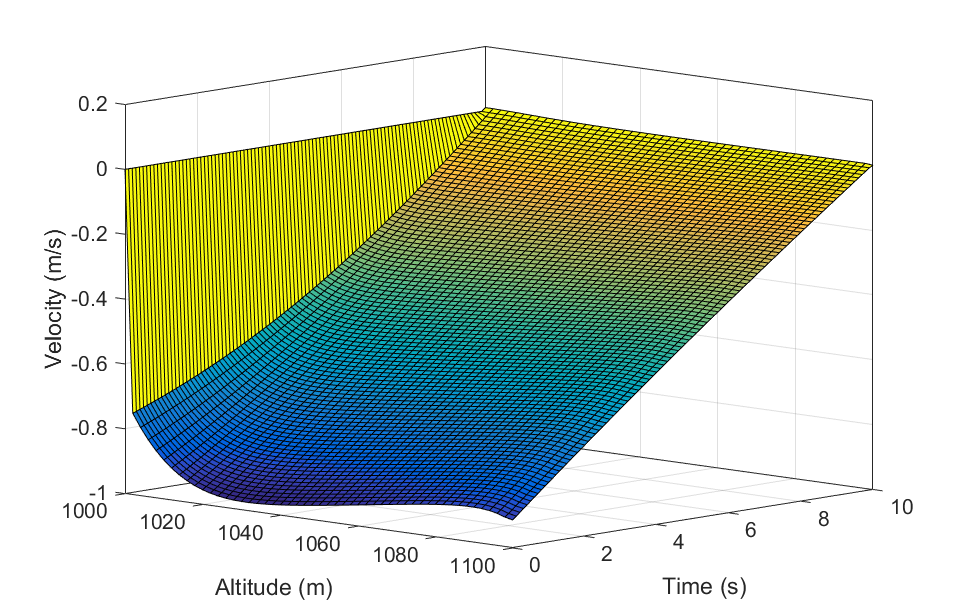}
 
 				\caption{Optimal velocity control policy with uniform initial altitude distribution.}

 	\end{center}
 
 \end{figure}
which cause larger interference. Hence, to improve the SINR, each DSC need larger velocities, which cost more energy.

In Fig. 4, we show the optimal velocity policies for the DSCs. Each DSC can adjust the dynamical system to catch the corresponding velocity calculated at the beginning of each distributed control time interval based on its current altitude at each time instant. It is notable that the length of the control time interval can be set flexibly. All DSCs drop with different velocity after the start. As illustrated in this figure, to improve the SINR, all DSCs tend to drop to shorten the distance to the user. And All DSCs' velocities decrease with time. That is because, as all DSCs have the downward tend, which result in the increased interference, each DSCs’ altitude reduction make lower impacts on the value function.

%



At last, we investigate the average spectrum effectiveness of the DSCs during the control period. We refer to the performence index average SINR to describe the improvement in quality of communications from our proposed algorithm. For comparision, we derive a uniform velocity control scheme as a benchmark. We assume a proper descent velocity for all DSCs, in case it costs much more power for movement. And we also consider the static strategy. The simulation results are shown in Fig. 5. The two curves represent the average SINR of all DSCs forward in time. The results illustrate that in comparison to the static strategy and the uniform velocity scheme, the control policy obtained from the proposed algorithm has a better performence.

 \begin{figure}[ht]
 	\begin{center}
 
 		\includegraphics[scale=0.30]{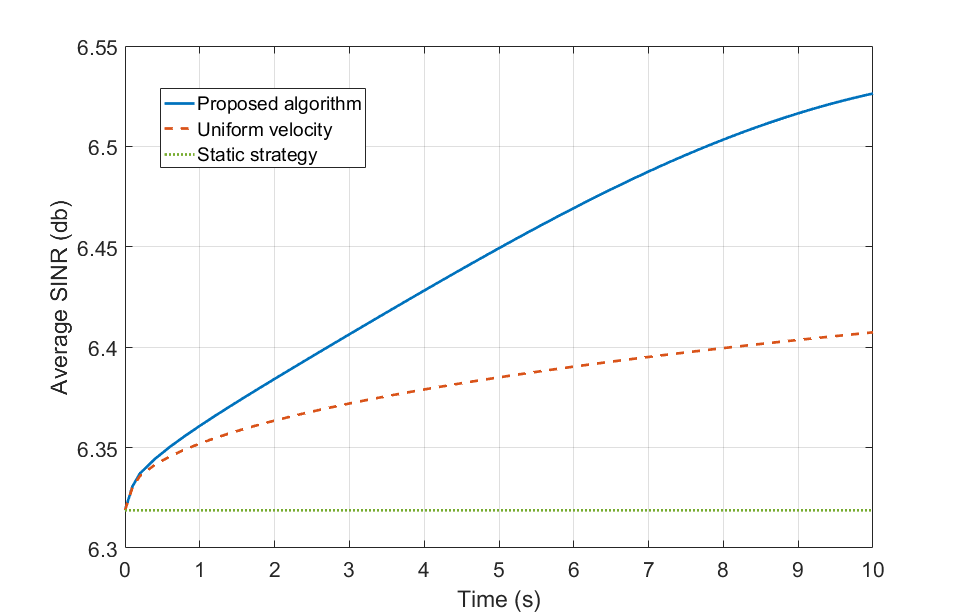}
 
 				\caption{Variation of average SINR at a generic user with time.}

 	\end{center}
 
 \end{figure}

\vspace{0.2cm}
\section{Conclusion}\label{section_6}
\vspace{0.2cm}
In this paper, we investigate an altitude control problem of the interference-aware DSCs. In this problem we take into account of both the motive consumption during the predefined period and the instantaneous communication quality. In the context of the high density and separate deployments of DSCs, a distributed altitude control method based on the MFG theoretic framework is proposed. Assuming that all DSCs have the same transmit power, we apply a MFA to design the cost function. Based on the derived coupled HJB and FPK partial differential equations, the solutions of the MFG are obtained. An iterative finite difference technique is proposed to solve the mean field equations based on the upwind scheme. The numerical results are presented to show the distributions and behaviors of DSCs. Also, the performence of the proposed optimal velocity control algorithm is illstrated compared with some benchmarks. One of the advantages of this algorithm is that, in practice, the algorithm can be executed distributively, which can lessen the burden on the center managers.

\vspace{0.5cm}
\bibliographystyle{IEEEtran}
\bibliography{bibfile_gen}

\end{document}